\documentclass[conference]{IEEEtran}

\usepackage{cite}
\usepackage{dblfloatfix} 
\usepackage{amsmath,amssymb,amsfonts}
\usepackage{algorithmic}
\usepackage{graphicx}
\usepackage{textcomp}
\usepackage{xcolor}
\usepackage{circuitikz}
\def\BibTeX{{\rm B\kern-.05em{\sc i\kern-.025em b}\kern-.08em
    T\kern-.1667em\lower.7ex\hbox{E}\kern-.125emX}}

\usepackage{hyperref}
\usepackage{siunitx}

\begin{document}

\title{Data-Driven Thermal Modelling for Anomaly Detection in Electric Vehicle Charging Stations}

\author{\IEEEauthorblockN{Pere Izquierdo G\'omez, Alberto Barragan Moreno, Jun Lin, Tomislav Dragi\v{c}evi\'{c}}
\IEEEauthorblockN{Department of Wind and Energy Systems, Technical University of Denmark, Kgs. Lyngby, 2800, Denmark}
\IEEEauthorblockA{\{pizgo, abamo, junli, tomdr\}@dtu.dk}

\thanks{P. I. Gómez, A. B. Moreno, J. Lin, and T. Dragi\v{c}evi\'{c} are with the Department
of Electrical Engineering, Technical University of Denmark, Kgs. Lyngby,
2800, Denmark. E-mail of correspondent author: pizgo@dtu.dk.}
}

\maketitle

\begin{abstract}
The rapid growth of the electric vehicle (EV) sector is giving rise to many infrastructural challenges. One such challenge is its requirement for the widespread development of EV charging stations which must be able to provide large amounts of power in an on-demand basis. This can cause large stresses on the electrical and electronic components of the charging infrastructure---negatively affecting its reliability as well as leading to increased maintenance and operation costs. This paper proposes a human-interpretable data-driven method for anomaly detection in EV charging stations, aiming to provide information for the condition monitoring and predictive maintenance of power converters within such a station. To this end, a model of a high-efficiency EV charging station is used to simulate the thermal behaviour of EV charger power converter modules, creating a data set for the training of neural network models. These machine learning models are then employed for the identification of anomalous performance.  
\end{abstract}


\section{Introduction}
The transportation sector is a main contributor to global greenhouse gas emissions, as it is the consumer of an estimated 65.2\% of the world's oil products \cite{b1}. Therefore, the electrification of the transportation sector plays a key role in the transition towards a decarbonised global society. With transport-related CO\textsubscript{2} emissions on a globally increasing trend \cite{b2}, this electrification process is becoming increasingly urgent.

One of the main tools to undertake this transition is the widespread replacement of internal combustion engine vehicles (ICEV) with electric vehicles (EVs), which are receiving ample support from policy makers. In an attempt to accelerate the adoption of EVs, many of the world's major economies currently offer fiscal incentives to purchasers, which have been instrumental in the recent growth of the sector. In 2020, despite the COVID-19 pandemic leading to a 16\% decrease in the sale of all cars when compared to the previous year, EV sales increased by 40\% \cite{b3}. This growth entails increased requirements of public charging infrastructure, which aim to fulfill a similar role as that of gas stations for ICEVs. Illustrating this fact, the number of charging ports in the United States grew from approximately nearly 34,000 to more than 85,000 between the years 2016 and 2020 \cite{b4}. 

For EV charging infrastructure development to adapt to the growth of EV fleets, charging station economic viability should be maximized. A major source of costs in EV charging station projects, which should therefore be minimized, is that of maintenance and downtime. In a 2012 paper, A. Schroeder and T. Traber estimated maintenance and repair costs at up to 10\% of material costs per year \cite{b5}, while in 2016, J. Serradilla, J. Wardle, P. Blythe, and J. Gibbon estimated planned and unplanned maintenance costs at 3\% and 4\% of charger cost per year, respectively \cite{b6}. One way to minimize maintenance costs is through the use of predictive maintenance, which makes use of condition monitoring techniques to optimize the scheduling of maintenance operations. Through the use of such techniques, wear-out failures can sometimes be identified before compromising intended operation, and some catastrophic failures can be prevented. 

Anomaly detection---which includes outlier and novelty detection---is one of the methods that can be employed to extract useful information about the condition of a given system. At its core, anomaly detection consists of the identification of data samples that fall outside of a given probability distribution \cite{b7}\cite{b8}. By identifying this distribution with a model of a given system, anomaly detection can be used to identify observations that do not correspond to the expected behaviour of the system. Machine learning techniques are therefore a natural fit to anomaly detection, as they can be used to obtain arbitrarily complex models to reflect the behavior of the system in an accurate manner. In particular, deep neural networks, due to their universal function approximation properties \cite{b9}, have found ample success in system identification and modelling \cite{b10}\cite{b11}.

This paper proposes a method for data-driven anomaly detection in EV charging stations, based on the modelling of the thermal behavior of individual power converters through machine learning models. The proposed method is based on the following assumptions:
\begin{itemize}
    \item Power converters are the most critical part of EV charging stations, and power semiconductor devices are their most fragile components \cite{b12}.
    
    \item Temperature stress is one of the main causes of the destruction of power electronic components, and maximum steady-state temperature is one of the two halves of this type of stress, with the other being temperature cycling \cite{b13}.
\end{itemize}

Therefore, monitoring the temperature of power electronic components, either directly (e.g. measuring switching device junction temperature) or indirectly (e.g. measuring heat sink temperature), can provide valuable information on the status of the converter's cooling system, its relative usage rate, or degradation of the heat sink and/or device materials. By using this information to schedule maintenance tasks, wear-out and catastrophic failures can be mitigated or prevented altogether.

\section{The Charging Station Model}
\label{sec:charging_station_model}

\subsection{General structure}
\begin{figure}[t]
    \centering
    \includegraphics[width=\columnwidth]{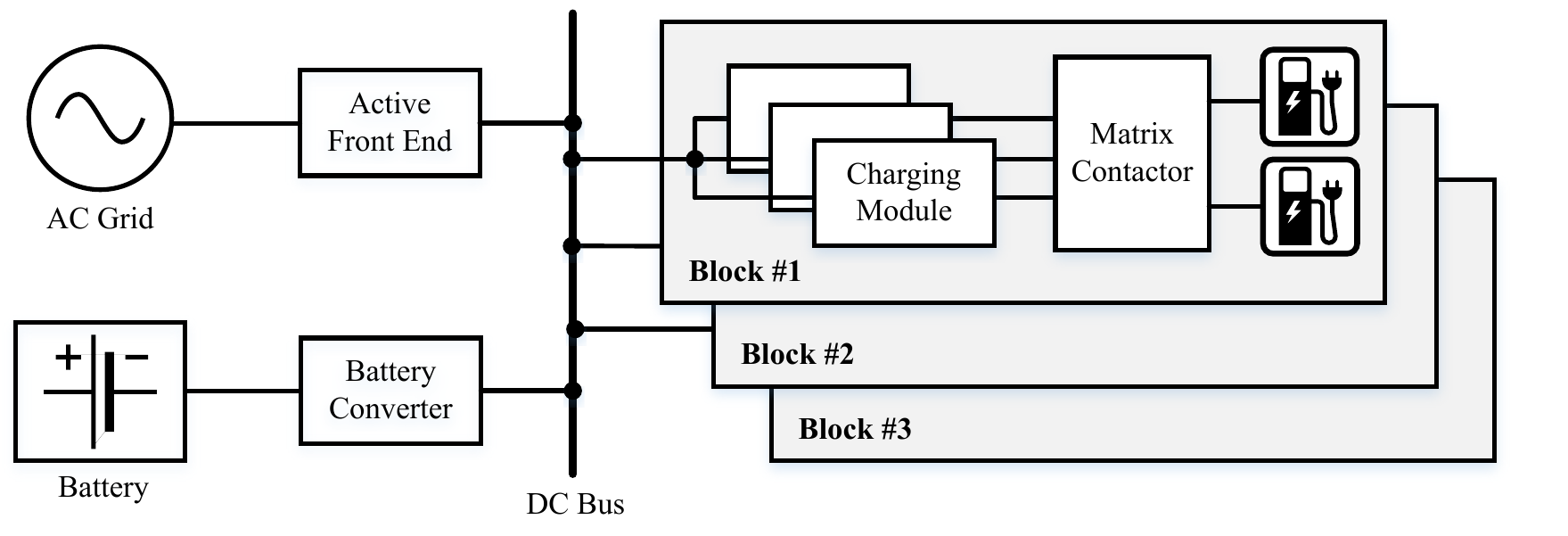}
    \caption{Block diagram representing the EV charging station topology under study.}
    \label{fig:ConverterBlockDiagram}
\end{figure}

The study is based on a particular topology of EV charging station, a block diagram of which is shown in Fig.~\ref{fig:ConverterBlockDiagram}. In this topology, an active front end (controlled rectifier) is used to interface the ac grid with a common dc bus, and a stationary battery is connected to this same dc bus through a dc-to-dc converter. The topology is then structured into three blocks connected to the dc bus, each consisting of three charging module dc-to-dc converters, which feed two charging posts through a matrix contactor. Such a modular design provides the advantages of being easily expandable---by simply connecting more blocks to the dc bus---, and of allowing for converter efficiency optimization through the online selection of enabled charging module converters \cite{b14}.

\subsection{Load profiles}
Load profiles for each of the charging modules are obtained from two stochastic profiles: the charging curve of each EV, and the probability distribution of charging events over the course of a day. 

In the developed simulation, EV charging profiles are considered to follow a two-stage profile of constant current charging followed by constant voltage charging. These profiles are defined by five parameters: initial state of charge (SoC), initial constant power, final SoC for constant power charging, exponential decay factor, and final SoC \cite{b15}. For each EV being charged at the station, each of these parameters is sampled from a normal probability distribution, with e.g. peak charging power having a 150 kW mean and a 20 kW standard deviation, and EV battery capacity having an average of 90 kWh and a standard deviation of 10 kWh.

The modelling of charging event probability is based on data published in \cite{b16} and \cite{b17}. Charging events are sampled from the obtained distribution, and load profiles are obtained based on a first-come first-served policy \cite{b18}.

\subsection{Thermal model}
\begin{figure}[t]
\begin{center}
\ctikzset{bipoles/length=1.2cm}
\begin{circuitikz}[european, scale = 0.8, transform shape]
    \draw (0, 0) node[tlground]{} -- (0, 2)
    to [american current source, l=$P_{loss}$] (1.5, 2)
    to [R=$R_{eq}$] (3.5, 2) -- (4, 2)
    to [C=$C_{hs}$] (4, 0) node[tlground]{};
    \draw (4, 2) to [R=$R_{hs}$] (7, 2)
    to [american voltage source, l=$T_{amb}$] (7, 0) node[tlground]{};
    \draw (4, 2) node[circ, label=$T_{hs}$]{};
\end{circuitikz}
\caption{Circuit diagram representing the thermal model applied to the power loss of each charging module.}
\label{fig:ThermalModel}
\end{center}
\end{figure}
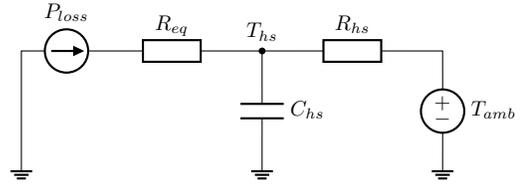

Using the requested loads at each charging post, an online optimization algorithm assigns active charging modules (sized at 60 kW) to maximize global efficiency \cite{b19}. The efficiency map of each module is adapted from \cite{b20}, and allows us to estimate power losses based on the assigned requested power. Module losses are then used as inputs to a thermal model, represented in Fig.~\ref{fig:ThermalModel}, where $P_{loss}$ is the power loss of the module, $R_{eq}$ is an equivalent thermal resistance for the semiconductor devices, $T_{hs}$ is the heat sink temperature, $R_{hs}$ and $C_{hs}$ are, respectively, the thermal resistance and capacitance of the heat sink, and $T_{amb}$ is the ambient temperature.

$R_{eq}$, $R_{hs}$, and $C_{hs}$ are sampled from normal distributions for each of the charging modules, with $\bar{R}_{eq} = \SI{1}{mK/W}$, $\bar{R}_{hs} = \SI{1.5}{mK/W}$, and the time constant of heat sink $RC$ branch at $\bar{\tau}_{hs} = \SI{120}{s}$. The standard deviation of each parameter is set at 5\% of its mean value. Ambient temperature $T_{amb}$ is assumed constant at $\SI{20}{\celsius}$.

\begin{figure}[b]
    \centering
    \includegraphics[width=\columnwidth]{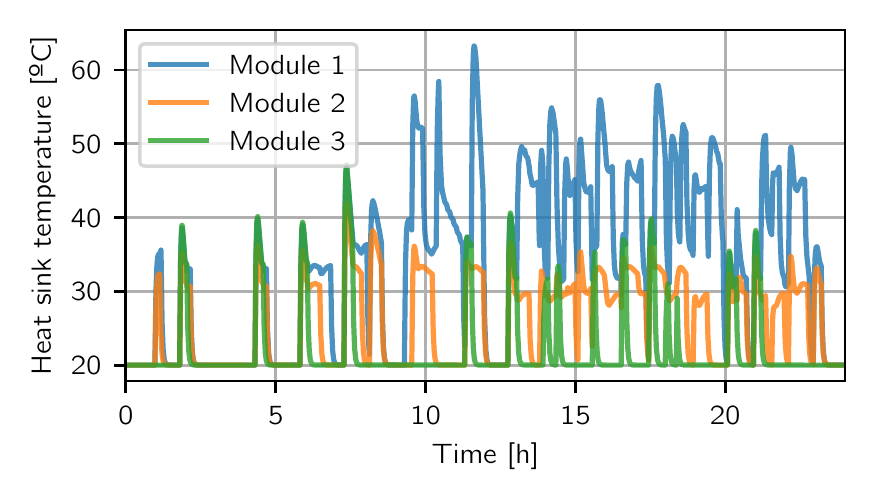}
    \caption{Simulated heat sink temperature profiles of the three converter modules corresponding to the first block of the charging station. Module 1 is active more frequently than module 2, which is in turn more active than module 3. This reflects directly on their simulated heat sink temperatures.}
    \label{fig:TemperatureProfiles}
\end{figure}

By joining these simulations, we obtain heat sink temperature profiles for each of the charging modules of the station. The final simulation samples 24-hour periods at a rate of 7.2 s, recording the estimated power loss and heat sink temperature corresponding to each of the nine charging modules. The heat sink temperature profiles corresponding to the first three modules are shown in Fig.~\ref{fig:TemperatureProfiles}.

\begin{figure*}[t]
    \centering
    \includegraphics[width=\textwidth]{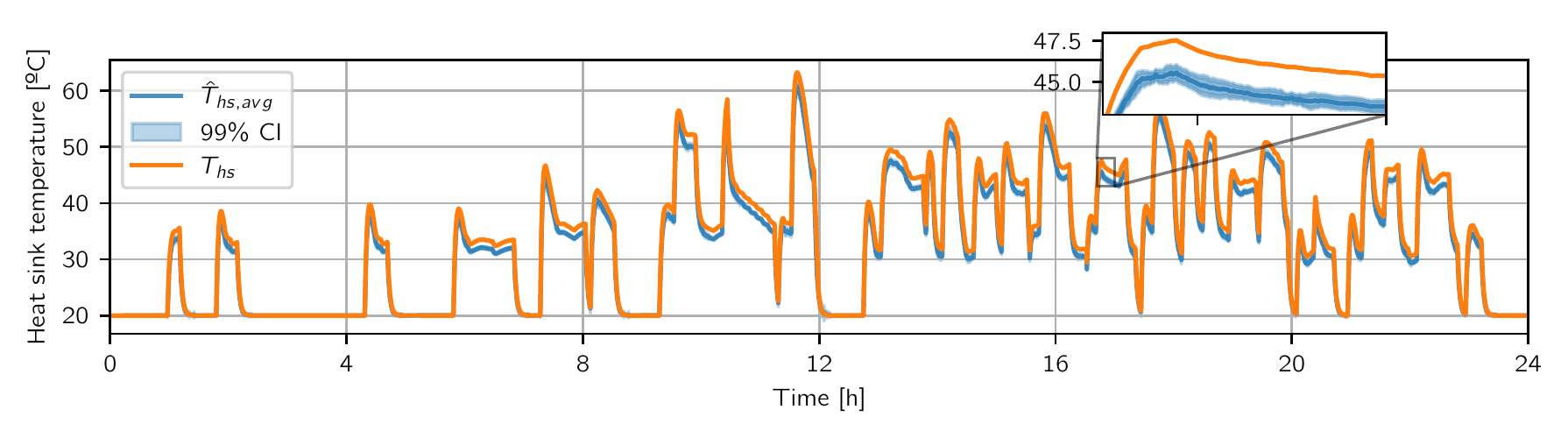}
    \caption{Predictions of the ensemble of models on the training data set corresponding to the charging module 1. The plot includes the average prediction of heat sink temperature $\hat{T}_{hs, avg}$, its corresponding 99\% CI constructed with a \textit{t}-statistic, and the ground truth for heat sink temperature $T_{hs}$.}
    \label{fig:TrainingPredictions}
\end{figure*}

The figure illustrates how the profiles within a block of modules differ significantly from one another, as the simulation prioritizes loading the first module(s). This leads to higher losses for the first---and to a lesser extend, the second---module and, in turn, higher heat sink temperatures. The figure also shows how, in the described simulation, charging activity is more prominent during daytime hours.

\section{Data-Driven Modelling of Thermal Profiles}
The generated data set is then used for the training of neural network models, with the goal of estimating the heat sink temperature profiles for a given module power loss profile, during non-anomalous operation. 

The chosen model architecture, obtained through experimentation, is an ensemble of 10 neural networks, where each one predicts the current heat sink temperature based on a window of the last 125 power loss estimates (for a single charging module), which in turn correspond to 15 minutes of operation. Each neural network is feedforward, fully-connected, and contains two hidden layers of 128 and 64 units. The prediction for the $i$-th neural network can therefore be obtained as:
\begin{align}
    \mathbf{x}_{1,i} &= \phi(\mathbf{W}_{1,i} \mathbf{p_{loss}} + \mathbf{b}_{1,i}) \\
    \mathbf{x}_{2,i} &= \phi(\mathbf{W}_{2,i} \mathbf{x}_{1,i} + \mathbf{b}_{2,i}) \\
    \hat{T}_{hs,i} &= \mathbf{W}_{3,i} \mathbf{x}_{2,i} + b_{3,i}
\end{align}
Where $\mathbf{p_{loss}}$ is a vector of module power losses, $\mathbf{x}_{j,i}$ is a vector of activation values of the $j$-th layer, $\mathbf{W}_{j,i}$ is a matrix of weights of the $j$-th layer, $\mathbf{b}_{j,i}$ is a vector of biases of the $j$-th layer, and $\hat{T}_{hs, i}$ is the predicted heat sink temperature. $\phi()$ is a nonlinear activation function, in this implementation, the rectified linear unit (ReLU) function. Ensemble mean predictions are then obtained by averaging $\mathbf{\hat{T}_{hs}}$ over $i$.

The use of an ensemble of neural networks, when compared to a single neural network, improves prediction robustness. Moreover, it allows for the computation of additional metrics, such as prediction variance, which can in turn be used to estimate confidence intervals. 

Each model in the ensemble is trained sequentially, using a mean squared error loss function, the backpropagation algorithm and the Adam optimizer \cite{b21}. The training data set, corresponding to a 24-hour simulation (for a total of about 100k data points), is randomly split into 80\%-20\% for training and validation, respectively. Each model is then trained for 100 epochs in batches of 64 samples, keeping the model corresponding to the iteration with the lowest validation loss. The implementation of the models and their training is based on the PyTorch library \cite{b22}.

\section{Results and Discussion}

\subsection{Training results}
To illustrate the performance of the trained model on the training data set, Fig.~\ref{fig:TrainingPredictions} shows the heat sink temperature ground truth $T_{hs}$, the average prediction of the model ensemble $\hat{T}_{hs, avg}$, and the area corresponding to a 99\% confidence interval (CI), for the first charging module (also shown in Fig.~\ref{fig:TemperatureProfiles}). The confidence intervals are constructed for each sample using the standard deviation of model predictions and the \textit{t}-statistic corresponding to the desired confidence level.

From Fig.~\ref{fig:TrainingPredictions}, it is clear that the trained model consistently predicts temperatures at values up to $\SI{3}{\celsius}$ lower than the ground truth. Moreover, ground truth values consistently fall outside of the confidence interval. This is, however, the expected behaviour of the model. The bias in the predictions can be explained by the fact that each model in the ensemble is trained to predict the \textit{average} expected heat sink temperature; since the thermal model parameters are stochastic, the heat sink temperatures of some modules are expected to differ from the average behaviour. Indeed, for this module, the values of $R_{eq}$ and $R_{hs}$ are both higher than their distribution mean. Regarding confidence intervals, when the individual models obtain similar predictions for a given set of inputs, the confidence interval will be accordingly small. Their use is therefore not in quantifying the variability of temperatures across different models, but rather in indicating the prediction uncertainty. Input samples that differ significantly from training data will cause larger differences in individual model predictions, and therefore confidence intervals will be larger for samples where predictions are less certain.

\subsection{Anomaly metrics}
In order to analyze whether the thermal behavior of a charging module can be classified as anomalous, several metrics can be defined based on prediction errors. A basic metric is the absolute error (AE) between the mean prediction of the ensemble and the ground truth:

\begin{equation}
    AE = | T_{hs}-\hat{T}_{hs, avg} |
\end{equation}

Although other metrics, such as Mahalanobis distance, have been successfully applied in anomaly detection methods \cite{b22}, absolute error metrics perform best in this problem. AE can then be normalized using the sample standard deviation $s$ of the predictions to obtain a metric more resilient to prediction uncertainty:
\begin{gather}
    s = \sqrt{\frac{1}{N-1} \sum^N_{i=1}(\hat{T}_{hs, i} - \hat{T}_{hs, avg})^2} \\
    AE_{norm} = \frac{| T_{hs}-\hat{T}_{hs, avg} |}{s}
\end{gather}
\\
Where $N$ is the number of model predictions for each sample---in this case, 10. Absolute error metrics are processed through moving average filters to improve their stability, allowing for more reliable comparisons. Three different types of moving average are considered: the simple moving average (SMA), the cumulative moving average (CMA), and the exponential moving average (EMA). These are defined as follows:

\begin{align}
    SMA_{k, n} &= \frac{AE_{k} + AE_{k-1} + \cdots + AE_{k-n+1}}{n}  \\
    CMA_{k} &= \frac{AE_{k} + AE_{k-1} + \cdots + AE_{1}}{k} \\
    EMA_{k} &= 
    \begin{cases}
        AE_1, & k = 1 \\
        \alpha \cdot AE_k + (1 - \alpha) \cdot EMA_{k-1}, & k > 1
    \end{cases}
\end{align}
\\
Where $k$ is the current period of the sample, $n$ is the number of samples considered in the moving window of the simple moving average, and $\alpha$ is the decay rate of the exponential moving average. The absolute error $AE_k$ may be either raw or normalized.

\begin{figure*}[t]
    \centering
    \includegraphics[width=\textwidth]{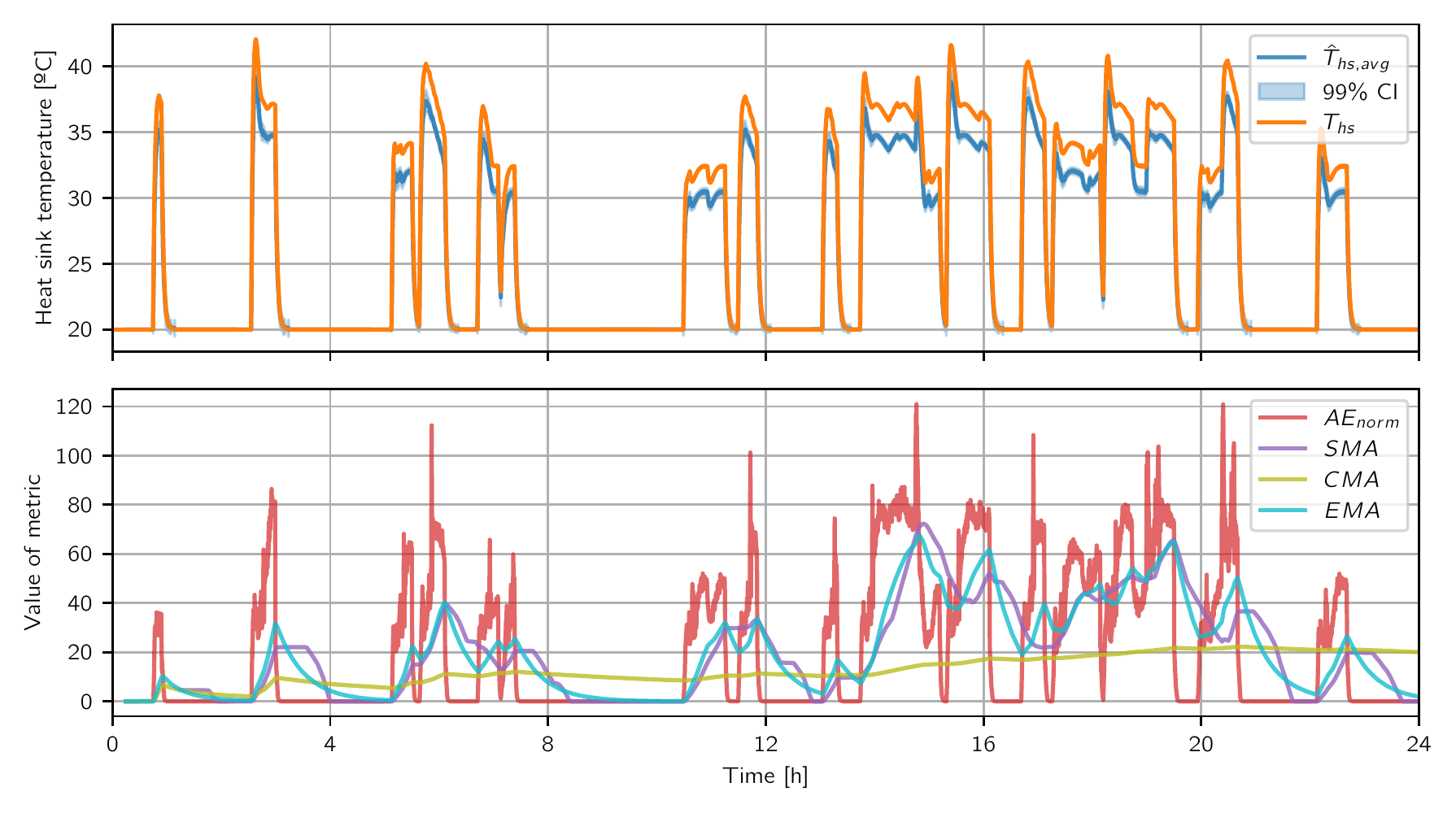}
    \caption{Prediction results corresponding to an anomalous charging module. The top graph shows the predicted heat sink temperature $\hat{T}_{hs, avg}$, its 99\% confidence interval, and the ground truth for heat sink temperature $T_{hs}$. The bottom graph contains the corresponding anomaly metrics: normalized absolute error $AE_{norm}$, its simple moving average $SMA$ with a 1-hour rolling window, its cumulative moving average $CMA$, and its exponential moving average $EMA$ with $\alpha = \num{4e-3}$.}
    \label{fig:pred_results}
\end{figure*}

\subsection{Anomaly detection results}
In order to evaluate the performance of the method and the defined metrics, we generate a testing data set containing an anomalous charging module. To this end, the 24-hour simulation defined in section \ref{sec:charging_station_model} is repeated with new load profiles. Thermal model parameters are kept at the same values, with the sole exception of a 20\% increase in the value of $R_{hs}$ for the second charging module of the second block (i.e. the fifth module). A similar change could occur in a real application due to several factors, such as degradation of the thermal interface material of the heat sink, or a fault in the cooling fan of the system. The proposed method is henceforth evaluated on its ability to accurately identify this anomalous behaviour.

The model ensemble is thus used to make predictions on this data set, and the results corresponding to the fifth module are shown in Fig.~\ref{fig:pred_results}. The figure includes the model predictions in its upper graph, while its bottom graph displays the obtained values for the previously defined anomaly metrics. The three displayed moving averages $SMA$, $CMA$, and $EMA$ are all calculated based on the normalized version of the absolute prediction error. When comparing the prediction results obtained in Fig.~\ref{fig:TrainingPredictions} and Fig. \ref{fig:pred_results} it is not immediately clear whether the second one may qualify as anomalous while the first one may not, since temperatures remain consistently lower in Fig. \ref{fig:pred_results} due to the lower utilization of their corresponding module.

Fig.~\ref{fig:pred_results} shows the noisiness of the absolute error values $AE_{norm}$, which justifies employing moving averages for the comparison of prediction errors. The three displayed types of moving average have advantages and disadvantages with respect to one another. $CMA$ is stable but more influenced by inactive periods, where the temperature is stable and the error very low. $SMA$ and $EMA$ show similar results, but $SMA$ introduces more significant delays. For this reason, $EMA$ is selected as the metric for the final evaluation. 

\begin{figure}[t]
    \centering
    \includegraphics[width=\columnwidth]{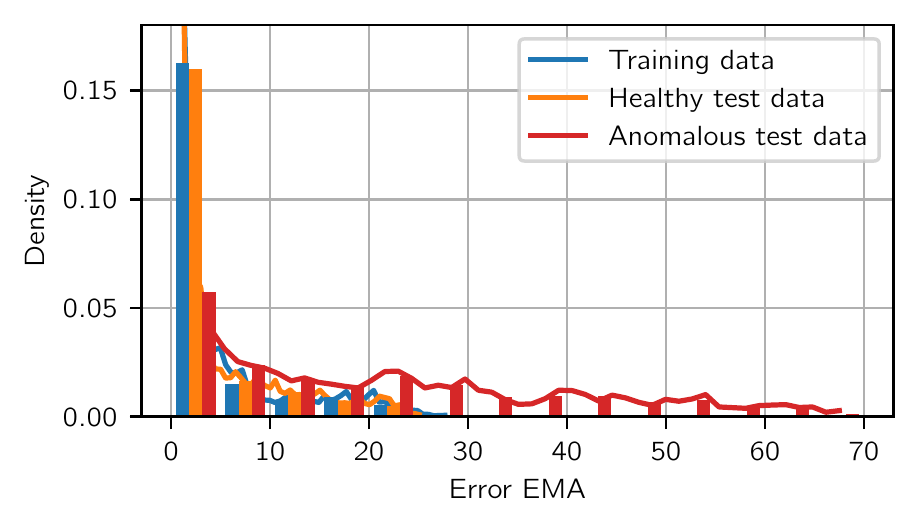}
    \caption{Histogram of the values of normalized absolute error $AE_{norm}$ processed with an exponential moving average filter with $\alpha = \num{4e-3}$. The values shown correspond to three separate data sets: training data, testing data excluding the anomalous module (healthy data) and testing data corresponding to the anomalous module.}
    \label{fig:histogram}
\end{figure}

To this end, the same procedure is applied to the complete training data set and to the complete testing data set. Values for the $EMA$ of $AE_{norm}$ are recorded and plotted in the histogram shown in Fig.~\ref{fig:histogram}. The figure clearly shows how the distribution of processed errors is very similar for the training and the healthy test data sets, which indicates that the model has been adequately fit to the training data. More importantly, the distribution of processed errors in the anomalous test data is significantly different than that of the others, with consistently greater values. Specifically, over 25\% of anomalous data samples are assigned a metric above 30. 

We may therefore state the decision rule for the classification of anomalous behavior as finding over 20\% of $AE_{norm}$-$EMA$ values above 30 throughout a 24-hour period.

\section{Conclusions}

This article proposes a method for anomaly detection in a population of power electronic converter devices, within the context of a modular EV charging station. The method is based on the following steps:
\begin{enumerate}
    \item Obtain a data set corresponding to the healthy operation of the devices under study, including estimates or measurements of heat sink temperatures and power losses.
    
    \item Use the data set to train one or several machine learning models. In the presented case, the model takes the form of an ensemble of feedforward, fully-connected neural networks.
    
    \item Deploy the model(s), obtaining real-time heat sink temperature predictions and their associated error metrics. These may take the form of absolute error and be filtered using moving averages.
    
    \item Compare the values of the error metric(s) between the different devices, over a certain period of time. Devices that consistently reach significantly greater values may be concluded to present anomalous behaviour.
\end{enumerate}

For the presented case study, the combination of an absolute error metric an an exponential moving average filter have resulted in more reliable anomaly detection than more advanced metrics such as the Mahalanobis distance. Other types of filters, such as the simple moving average, perform very similarly to the exponential moving average, given that their parameters are tuned appropriately.

The method has been shown to perform accurately and reliably on simulation data, although further work is required to verify its capabilities in a more realistic scenario.

\end{document}